\documentclass[aps,prc,showpacs,twocolumn,superscriptaddress,floatfix]{revtex4}

\usepackage[dvips]{graphicx}   % Include figure files

\begin{document}

\title{Centrality Dependence of Chemical Freeze-out in Au+Au Collisions at RHIC}

\author{Masashi Kaneta}
  \affiliation{RIKEN-BNL Research Center, Brookhaven National Laboratory, Upton, New York 11973}
\author{Nu Xu}
  \affiliation{Lawrence Berkeley National Laboratory, Berkeley, California 94720}
\noaffiliation

\date{\today}

\begin{abstract}
  This is a write-up of a poster presented in the international conference Quark Matter 2004 at Oakland, CA. 
  We will report centrality dependence of chemical freeze-out temperature ($T_{ch}$), light quark chemical potential ($\mu_{q}$), strange quark chemical potential ($\mu_{s}$), and strangeness saturation factor ($\gamma_{s}$) in Au+Au collisions at $\sqrt{s_{NN}}$ = 130 and 200 GeV.
  A systematic study for combination of ratios for chemical freeze-out fit is studied and we have found small combination dependences.
  The results show $\gamma_{s}$ increasing with centrality but the other parameters have less sensitivity to the centralities.
\end{abstract}


\maketitle

\section{INTRODUCTION}
  The statistical model approach is established by analysis of particle ratios of the high energy heavy ion collisions in GSI-SIS to CERN-SPS energy~\cite{PL_B262_1991_333,PL_B344_1995_43,ZP_C74_1997_319,PL_B365_1996_1,PR_C53_1996_1353,JP_G25_1999_295,PR_C59_1999_1663} and elementary collisions ($e^{+}e^{-}$, $pp$, and $p\bar{p}$)~\cite{ZP_C69_1996_485,ZP_C76_1997_269}.
  The model describes the particle ratios by the chemical freeze-out temperature ($T_{ch}$), the chemical potential ($\mu$), and the strangeness saturation factor ($\gamma_{s}$).
  Many feature of the data imply that a large degree of the chemical equilibration may be reached both at AGS and SPS energies excepting strangeness hadrons.
  There are the four most important results.
  \begin{enumerate}
    \item
      At high energy collisions the chemical freeze-out (inelastic collisions cease) occurs at about 150-170 MeV and it is `universal' to both elementary and the heavy ion collisions;
    \item
      The strangeness is not fully equilibrated because $\gamma_{s}$ is 0.5-0.8~\cite{PR_C53_1996_1353,ZP_C69_1996_485,ZP_C76_1997_269} (if strangeness is in equilibration,  $\gamma_{s}$=1~\cite{PL_B262_1991_333});
    \item
      The kinetic freeze-out (elastic scatterings cease) occurs at a lower temperature $\sim$ 100-120 MeV;
    \item
      The compilation of the freeze-out parameters~\cite{PRL_81_1998_5284} in the heavy ion collisions in the energy range from 1 - 200 A$\cdot$GeV shows that a constant energy per particle $\langle E \rangle / \langle N \rangle \sim 1$ GeV can reproduce the behavior in the temperature-potential ($T_{ch} - \mu_{_B}$) plane~\cite{PRL_81_1998_5284}.
  \end{enumerate}
 We have many hadron yields and ratios including multi-strange hadrons as a function of centrality in Au+Au collisions at $\sqrt{s_{NN}}$ = 130 and 200 GeV.
 They allow us to study centrality dependence of chemical freeze-out at RHIC energy.

\section{MODEL}
  We adopt the grand canonical ensemble for a chemical freeze-out model based on Ref.~\cite{PR_C59_1999_1637} and the model used in Ref.~\cite{JP_G27_2001_589,NP_A698_2002_306,PR_C66_2002_044907}.
  We assume no difference between $u$ and $d$ quark chemical potentials.
  Therefore, the hadron gas is described by a chemical freeze-out temperature ($T_{ch}$), light ($u$ and $d$) quark potential ($\mu_q$), strange quark potential ($\mu_s$), and strangeness saturation factor ($\gamma_s$).
  The density of a particle $i$ in the hadron gas is given by:
  \begin{equation}
    \label{eqn:rho}
    \rho_{i} = {\gamma_s}^{{\langle}s+\bar{s}{\rangle}_{i}} \frac{g_{i}}{2\pi^{2}} {T_{ch}}^3
               \left( \frac{m_i}{T_{ch}} \right)^2
               K_2(m_i/T_{ch}) \
               {\lambda_q}^{Q_i} \ {\lambda_s}^{s_i}
  \end{equation}
  where $m_i$ is the mass of the hadron $i$, $g_i$ is the number of spin-isospin degree-of-freedom, $K_2$ is the second-order modified Bessel function and,
  \[
    \lambda_q = \exp(\mu_q/T_{ch}),  \ \ \
    \lambda_s = \exp(\mu_s/T_{ch}).
  \]
  The potential $\mu_q$ is for u/d/$\bar{\text{u}}$/$\bar{\text{d}}$ quarks, and $\mu_s$ is for s/$\bar{\text{s}}$ quarks.
  The $\mu_q$ is a third of baryon chemical potential $\mu_{B}$.
  $Q_i$ ans $s_i$ correspond to the net number of valence u/d quarks ($Q_{i}={\langle}u-\bar{u}+d-\bar{d}{\rangle}_{i}$), and s quark ($s_{i}={\langle}s-\bar{s}{\rangle}_{i}$) of particle species $i$, respectively.
  The RHIC data we used are from mid-rapidity region, therefore we does not apply any conservation low and treat all of chemical freeze-out parameters as free parameters for the fit.
  The factor $\gamma_s$ ($0{\le}\gamma_s$) is introduced to take account of possible incomplete chemical equilibration for strange particles~\cite{PL_B262_1991_333}.
  $\gamma_s$=1 signifies full strangeness chemical equilibrium and $\gamma_s$$>$1 shows over strangeness saturation.
  The power factor of $\gamma_s$ is total number of s and $\bar{\mbox{s}}$ quarks in the particle.
  It is not clear whether $\gamma_s$ should be unity or not.
  On the other hand, calculations in Refs.~\cite{PR_C66_2002_044907,PR_C64_2001_024901,EPJ_C5_1998_143,JP_G25_1999_295} support $\gamma_s$$<$1.
  Therefore, we adopt $\gamma_s$ as a free parameter.
  Note that we use the Boltzmann approximation for all hadrons except pions, where the Bose distribution is applied.
  The particle densities are computed by Eq.~(\ref{eqn:rho}) for the hadron gas including known resonances up to mass of 1.7 GeV/$c^2$, and the effect of broad resonance mass widths is taken into account~\cite{PR_C59_1999_1637}.
  The resonance decay to lower mass hadrons after chemical freeze-out is also taken into account.

  There are technical caveats in above approaches.
  The application of some statistical model requires that the measurement is done over the whole phase-space, {\it i.e.}, with 4$\pi$ particle yields.
  This is because conservation laws that apply to the collisions are only valid for global measurements, not locally.
  In all experiments the coverage of the phase-space is limited.
  Therefore, extracting the yields measured within limited phase-space to 4$\pi$ yields is almost impossible, especially for the collider experiments.
  In addition, we should point out that the non-local effect is common to all global conservation laws (baryon number, charge, and so on), not only to the strangeness.

  We adopt the static model that assumes full phase space to limited phase space, that is, limited rapidity region.
  Ref.~\cite{JP_G25_1999_281} has studied effects of excluded volume correction, limited rapidity range, transverse flow, and longitudinal expansion to the ratios.
  Various effects which can severely distort the momentum spectra of the particles produced cancel out in such ratios~\cite{JP_G25_1999_281}.
  This addresses that the static model can be applied to the particle ratios in the limited rapidity range.

\section{ANALYSIS}
  Table~\ref{tbl:data} lists the centrality definition and references for the value of dN/dy and the ratios at mid-rapidity for each experiment.
  The centralities are defined by the independent detector and different acceptances among the experiments.
  The common parameter of the collision centrality is an average of number of participant ({${\langle}N_{part}{\rangle}$).
  Therefore, we use {${\langle}N_{part}\rangle$ to discuss the centrality dependence of the chemical freeze-out parameters.

  \begingroup
  \squeezetable
  \begin{table}
    \caption{\label{tbl:data}
             List of possible dN/dy and ratios from the experimental data in the Au+Au collisions at $\sqrt{s_{NN}}$ = 130 GeV and 200 GeV.
             The parenthetic particle name ($\langle{K^{*0}}\rangle$ and $\langle{K^{\pm}}\rangle$) means average of the particle and the anti-particle.
             The selection of event fraction is defined by
             (I) the multiplicity of charged particles from the Multiplicity Array~\cite{PL_B523_2001_227},
             (II) the energy deposit on Zero Degree Calorimeter (ZDC) and the number of charged particle measured by Beam-Beam Counter (BBC)~\cite{PRL_86_2001_3500},
             (III) the signal of the scintillator paddle counters~\cite{PRL85_2000_3100},
             (IV) the energy deposit on ZDC and the signal of Central Trigger Barrel (CTB)~\cite{nucl_ex_0111004},
             (V) the number of reconstructed charged tracks ($\eta<|0.75|$)~\cite{PRL_89_2002_092301},
             (VI) the number of reconstructed charged tracks ($\eta<|0.5|$).
             Additionally, some references show the number of participant and the $dN/\eta$ of negative charged hadrons for the event fraction.
             We marked them in the list by $\langle{N_{part}}\rangle$ and $dN_{h^-}/d\eta$.
            }
    \begin{ruledtabular}
      \begin{tabular}{lcccccc}
        $\sqrt{s_{NN}}$ & experiment & Refs.                                       & $dN/dy$ or ratio                                    & \multicolumn{3}{c}{centrality presented}                \\
        \hline                                                                                                                           
        130 GeV         & BRHAMS     & \cite{PRL_87_2001_112305}                   &  $\pi^{-}/\pi^{+}$                                  & (I)   & $\langle{N_{part}}\rangle$ &                    \\
                        &            & \cite{PRL_87_2001_112305}                   &  $\bar{p}/p$                                        & (I)   & $\langle{N_{part}}\rangle$ &                    \\
                        & PHENIX     & \cite{PRL_88_2002_242301,PRL_89_2002_092302}&  $\pi^{+}$, $\pi^{-}$                               & (II)  & $\langle{N_{part}}\rangle$ &                    \\
                        &            & \cite{PRL_88_2002_242301,PRL_89_2002_092302}&  $K^{+}$, $K^{-}$                                   & (II)  & $\langle{N_{part}}\rangle$ &                    \\
                        &            & \cite{PRL_88_2002_242301,PRL_89_2002_092302}&  $p$, $\bar{p}$                                     & (II)  & $\langle{N_{part}}\rangle$ &                    \\
                        &            & \cite{PRL_89_2002_092302}                   &  $\Lambda$, $\bar{\Lambda}$                         & (II)  & $\langle{N_{part}}\rangle$ &                    \\
                        & PHOBOS     & \cite{PRL_87_2002_102301}                   &  $\pi^{-}/\pi^{+}$                                  & (III) & $\langle{N_{part}}\rangle$ &                    \\
                        &            & \cite{PRL_87_2002_102301}                   &  $K^{-}/K^{+}$                                      & (III) & $\langle{N_{part}}\rangle$ &                    \\
                        &            & \cite{PRL_87_2002_102301}                   &  $\bar{p}/p$                                        & (III) & $\langle{N_{part}}\rangle$ &                    \\
                        & STAR       & \cite{nucl_ex_0111004}                      &  $\pi^-$                                            & (IV)  & $\langle{N_{part}}\rangle$ & $dN_{h^-}/d\eta$   \\
                        &            & \cite{nucl_ex_0206008}                      &  $K^{+}$, $K^{-}$, $K^0_S$                          & (V)   &                            & $dN_{h^-}/d\eta$   \\
                        &            & \cite{PR_C66_2002_061901}                   &  $\bar{K}^{*0}/K^{*0}$                              & (IV)  &                            &                    \\
                        &            & \cite{PR_C66_2002_061901}                   &  $\langle{K^{*0}}\rangle$                           & (IV)  &                            &                    \\
                        &            & \cite{PRL_87_2001_262302}                   &  $\bar{p}$                                          & (V)   &                            &                    \\
                        &            & \cite{PRL_86_2001_4778}                     &  $\bar{p}/p$                                        & (V)   &                            &                    \\
                        &            & \cite{PR_C65_2002_041901}                   &  $\phi$                                             & (V)   &                            &                    \\
                        &            & \cite{PRL_89_2002_092301}                   &  $\Lambda, \bar{\Lambda}$                           & (VI)  &                            &                    \\
                        &            & \cite{nucl_ex_0210032,nucl_ex_0307024}      &  $\Xi^{-}, \bar{\Xi}^+$                             & (V)   &                            &                    \\
                        &            & \cite{nucl_ex_0211017,nucl_ex_0307024}      &  $\Omega^{-} ,\bar{\Omega}^+$                       & (V)   &                            &                    \\
        200 GeV         & BRAHMS     & \cite{nucl_ex_0207006}                      &  $\pi^{-}/\pi^{+}$                                  & (I)   & $\langle{N_{part}}\rangle$ &                    \\
                        &            & \cite{nucl_ex_0207006}                      &  $K^{-}/K^{+}$                                      & (I)   & $\langle{N_{part}}\rangle$ &                    \\
                        &            & \cite{nucl_ex_0207006}                      &  $\bar{p}/p$                                        & (I)   & $\langle{N_{part}}\rangle$ &                    \\
                        & PHOBOS     & \cite{nucl_ex_0210037}                      &  $\pi^{-}/\pi^{+}$                                  & (III) & $\langle{N_{part}}\rangle$ &                    \\
                        &            & \cite{nucl_ex_0210037}                      &  $K^{-}/K^{+}$                                      & (III) & $\langle{N_{part}}\rangle$ &                    \\
                        &            & \cite{nucl_ex_0210037}                      &  $\bar{p}/p$                                        & (III) & $\langle{N_{part}}\rangle$ &                    \\
                        & PHENIX     & \cite{nucl_ex_0307022}                      &  $\pi^{+}$, $\pi^{-}$                               & (II)  & $\langle{N_{part}}\rangle$ &                    \\ 
                        &            & \cite{nucl_ex_0307022}                      &  $K^{+}$, $K^{-}$                                   & (II)  & $\langle{N_{part}}\rangle$ &                    \\ 
                        &            & \cite{nucl_ex_0307022}                      &  $p$, $\bar{p}$                                     & (II)  & $\langle{N_{part}}\rangle$ &                    \\ 
                        & STAR       & \cite{nucl_ex_0310004}                      &  $\pi^{+}$, $\pi^{-}$                               & (VI)  & $\langle{N_{part}}\rangle$ &                    \\
                        &            & \cite{nucl_ex_0310004}                      &  $K^{+}$, $K^{-}$                                   & (VI)  & $\langle{N_{part}}\rangle$ &                    \\
                        &            & \cite{nucl_ex_0310004}                      &  $p$, $\bar{p}$                                     & (VI)  & $\langle{N_{part}}\rangle$ &                    \\
                        &            & \cite{nucl_ex_0306034}                      &  $\langle{K^{*0}}\rangle$/$\langle{K^{\pm}}\rangle$ & (VI)  &                            &                    \\
                        &            & \cite{NP_A715_2003_466c}                    &  $\phi$                                             & (VI)  &                            &                    \\
                        &            & \cite{SQM2003_H_Long}                       &  $\Lambda + \bar{\Lambda}$                          & (VI)  &                            &                    \\
                        &            & \cite{HIC03_J_Castillo}                     &  $\Xi^{-}, \bar{\Xi}^{+}$                           & (VI)  &                            &                    \\
                        &            & \cite{nucl_ex_0211017}                      &  $\Omega^{-}, \bar{\Omega}^{+}$                     & (VI)  &                            &                     
      \end{tabular}
    \end{ruledtabular}
  \end{table}
  \endgroup

  We select two kind of centrality binning from STAR data.
  The first one is used for data set from charged pion, kaon and proton (centrality selection used in Ref.~\cite{nucl_ex_0206008,PRL_87_2001_262302} for 130 GeV and Ref.~\cite{nucl_ex_0310004} for 200 GeV).
  The second one has smaller number of bins but more particle ratios (centrality selection used in Ref.~\cite{PRL_89_2002_092301} for 130 GeV and Ref.~\cite{NP_A715_2003_466c} for 200 GeV).
  The other experiments have different centrality bins, therefore we need to estimate particle ratio for the centrality bins selected.
  At first, we interpolate the dN/dy of each particle assuming linear scaling of ${\langle}N_{part}{\rangle}$ between two data points.
  After we obtained dN/dy values for the ${\langle}N_{part}{\rangle}$, the particle ratios is computed from the dN/dy values.

  The chemical freeze-out model fit is applied to the particle ratios.
  A program package MINUIT~\cite{MINUIT} is used for the fit to find $\chi^{2}$ minimum points and the error of fit parameters is estimated as $\chi^{2}$ minimum + 1.
  The chemical freeze-out parameters seem to be sensitive to combination of particle ratios as discussed in Ref.~\cite{PR_C66_2002_044907}.
  Hence we check the following six combinations of particle ratios for the fit:
  (1) $\pi$, $K$, and $p$;
  (2) $\pi$, $K$, $p$ and $\Lambda$;
  (3) $\pi$, $K$, $p$, $\Lambda$,          $\phi$, and $\Xi$;
  (4) $\pi$, $K$, $p$, $\Lambda$, $K^{*}$, $\phi$, and $\Xi$;
  (5) $\pi$, $K$, $p$, $\Lambda$,          $\phi$, $\Xi$, and $\Omega$;
  (6) $\pi$, $K$, $p$, $\Lambda$, $K^{*}$, $\phi$, $\Xi$, and $\Omega$.

\section{RESULTS AND DISCUSSION}

  Figure~\ref{fig:chemical_parameter_vs_npart} shows centrality dependence of chemical freeze-out temperature $T_{ch}$,
  light-quark potential $\mu_{q}$ (= $\mu_{B}/3$ where $\mu_{B}$ is Baryonic chemical potential),
  strange quark potential $\mu_{s}$,
  and strangeness saturation factor $\gamma_{s}$
  as a function of ${\langle}N_{part}{\rangle}$ in 130 and 200 GeV Au+Au collisions.
  The markers correspond to combination of ratios used for fit as shown in the figure.
  Table \ref{tbl:chemical_freezeout_parameters} lists all of values shown in Fig.~\ref{fig:chemical_parameter_vs_npart}.

  The chemical freeze-out parameters vary with combination of particle ratios and shows larger variation in 130 GeV Au+Au than 200 GeV.
  Taking care the variation, $T_{ch}$, $\mu_{q}$, and $\mu_{s}$ seems to be constant with centrality in both 130 GeV and 200 GeV Au+Au collisions.
  The $\gamma_{s}$ is slightly increasing with centrality in 130 GeV when we focus the results by the fit to ratio from
  $\pi$, $K$, $p$, $\Lambda$, $K^{*}$, $\phi$, $\Xi$ (,and $\Omega$).
  On the other hand, we can see the rise of  $\gamma_{s}$ as a function of $\langle{N_{part}}\rangle$ in any combination of ratios in 200 GeV Au+Au.
  The $\gamma_{s}$ is about 0.5$\sim$0.7 in the heavy ion collisions at SPS energies~\cite{EPJ_C5_1998_143}, $e^{+}e^{-}$ at LEP, and $p\bar{p}$ at S$\bar{p}p$S~\cite{ZP_C69_1996_485,ZP_C76_1997_269}.
  Only Au+Au central collisions at RHIC energy ($\sqrt{s_{NN}}$ = 130 and 200 GeV) reach fully strangeness equilibration.
  The rise of $\gamma_{s}$ seems to be rapid until around ${\langle}N_{part}{\rangle}$ = 100-150 and then slowly increasing.

  \begin{figure}
    \includegraphics[width=8.5cm]{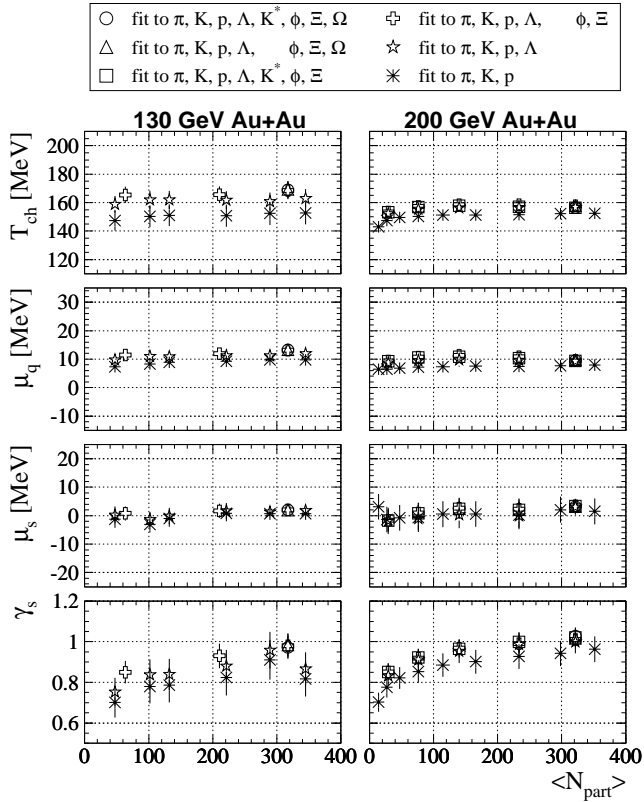}
    \caption{\label{fig:chemical_parameter_vs_npart}
             Chemical freeze-out temperature $T_{ch}$, light-quark potential $\mu_{q}$, strange quark potential $\mu_{s}$,
             and strangeness saturation factor $\gamma_{s}$ as a function of average number of participant.
             The different markers show different combination of particle ratio used for the fit.
            }
  \end{figure}

  \begin{figure}
    \includegraphics[width=8.5cm]{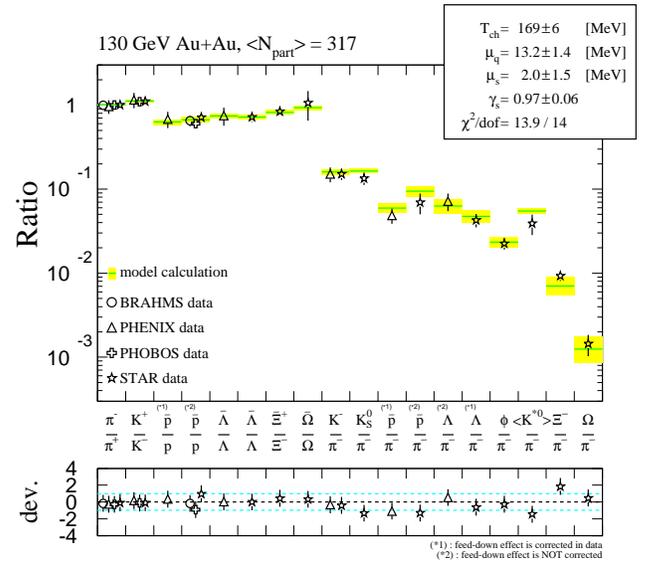}
    \caption{\label{fig:model_vs_data_130gev_w__all_mst_npart317}
             Comparison of fit results to the data in 130 GeV Au+Au central collisions (${\langle}N_{part}{\rangle}$=317).
             All of data from dataset (6) shown in the plot are used for the fit.
             The point with marker is experimental data.
             In the top plot, the model calculations of ratio from the chemical freeze-out parameters obtained are shown horizontal line.
             The variation in the error of the parameters is shown as shaded box.
             The bottom plot shows a deviation of data to the model, $(r_{exp}-r_{model})/\Delta{r_{exp}}$, where $r_{exp}$ is ratio from data, $r_{model}$ is ratio by model calculation, and  $\Delta{r_{exp}}$ is error of $r_{exp}$.
            }
  \end{figure}

  \begin{figure}
    \includegraphics[width=8.5cm]{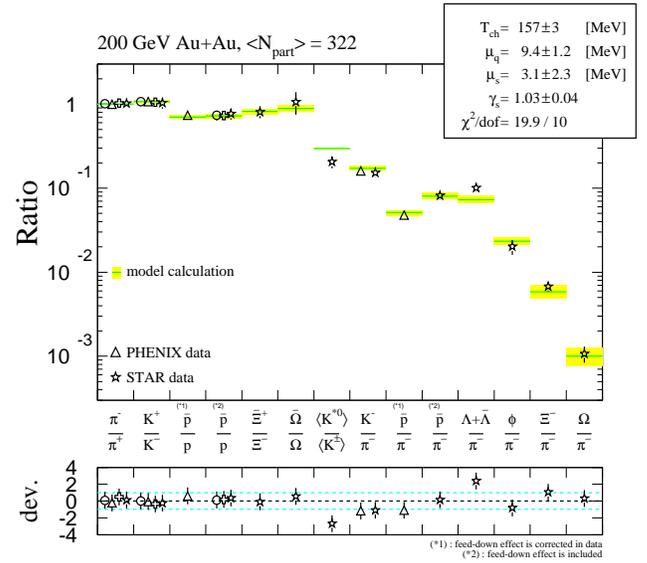}
    \caption{\label{fig:model_vs_data_200gev_w__all_mst_npart322}
             Comparison of fit results to the data in 200 GeV Au+Au central collisions (${\langle}N_{part}{\rangle}$=322).
             All of data from dataset (6) shown in the plot are used for the fit.
             The legend is the same with Fig.~\ref{fig:model_vs_data_130gev_w__all_mst_npart317}.
            }
  \end{figure}

  The $T_{ch}$, $\mu_{q}$, $\mu_{s}$ seem to be independent as a function of ${\langle}N_{part}{\rangle}$.
  Within errors and the variation of results from combination dependence,
  $T_{ch}$ is 150$-$170 MeV,
  $\mu_{q}$ is 8$-$11 MeV, and
  $\mu_{s}$ is -3$-$3 MeV.
  Small but finite $\mu_{q}$ reflects net baryon density is small but not zero around mid-rapidity.
  The $\mu_{s}$ is consistent with zero.
  It suggests that the chemical freeze-out occurs at phase boundary of a hadron gas.\cite{PL_B262_1991_333,PR_C37_1988_1452,PR_D51_1995_1086}.

  The chemical freeze-out parameter varies by data combination.
  In the central 130 GeV Au+Au data, $T_{ch}$, $\mu_{q}$, $\mu_{s}$, and $\gamma_{s}$ change about 10 MeV, 4MeV, 1MeV and 0.1, respectively.
  The deviation is small and consistent within error for all of parameters.
  The variation becomes smaller in the central 200 GeV Au+Au data.
  It is about 5 MeV, 2MeV, 2MeV and 0.06 for $T_{ch}$, $\mu_{q}$, $\mu_{s}$, and $\gamma_{s}$, respectively.
  The results from both energy shows that the $\gamma_{s}$ increases when we adopt multi-strangeness in the fit at a centrality bin.
  The $\gamma_{s}$ is sensitive to multi-strangeness, therefore the constrain changes stronger with increasing number of strangeness data point in the fit.
  However, it is not clear how the number of strangeness play the role in the constrain.
  We need to check this point with more data that will come soon from RHIC run4.

  Figure~\ref{fig:model_vs_data_130gev_w__all_mst_npart317} and \ref{fig:model_vs_data_200gev_w__all_mst_npart322} show a comparison of data to the model calculation.
  The data is included multi-strangeness and it is all of possible data sample at this time (dataset (6)).
  The model calculation shows good agreement in both energy, excepting $K^*$.
  In the 200GeV, the $(\Lambda+\bar{\Lambda})/\pi^-$ value from model also differ to the data.
  Note that the data $\Lambda+\bar{\Lambda}$ is obtained from a slide of the conference presentation from Ref.~\cite{SQM2003_H_Long} and the value seems to be changed in Quark Matter 2004 conference, therefore the results might be changed and we need to check in future.

\section{SUMMARY}

  We study centrality dependence of chemical freeze-out parameters in $\sqrt{s_{NN}}$ = 130 and 200 GeV Au+Au collisions.
  The $T_{ch}$, $\mu_{q}$, $\mu_{s}$ show independence as a function of ${\langle}N_{part}{\rangle}$.
  The most remarkable point is about centrality dependence of $\gamma_{s}$.
  The result shows $\gamma_{s}$ increasing with centrality and reach fully strangeness equilibration in only Au+Au central collisions at RHIC energy.
  The rise of $\gamma_{s}$ seems to be rapid until around ${\langle}N_{part}{\rangle}$ = 100-150 and then slowly increasing.
  The ratio combination dependence is also studied in both energy and we have found small combination dependences.
  The $T_{ch}$, $\mu_{q}$, $\mu_{s}$ seem to be independent as a function of ${\langle}N_{part}{\rangle}$.
  On the other hand, the $\gamma_{s}$ increases when we adopt multi-strangeness in the fit at a centrality bin.

  \begin{table*}
    \caption{\label{tbl:chemical_freezeout_parameters}
             Chemical freeze-out parameters for each average number of participant (${\langle}N_{part}{\rangle}$).
             The data set is related to combination of ratios used for the fit:
             (1) $\pi$, $K$, and $p$;
             (2) $\pi$, $K$, $p$ and $\Lambda$;
             (3) $\pi$, $K$, $p$, $\Lambda$,          $\phi$, and $\Xi$;
             (4) $\pi$, $K$, $p$, $\Lambda$, $K^{*}$, $\phi$, and $\Xi$;
             (5) $\pi$, $K$, $p$, $\Lambda$,          $\phi$, $\Xi$, and $\Omega$;
             (6) $\pi$, $K$, $p$, $\Lambda$, $K^{*}$, $\phi$, $\Xi$, and $\Omega$.
             The parameters are obtained by fit of chemical freeze-out model to the data.
             The error of parameters are defined as $\chi^2$ minimum + 1.
            }
    \begin{ruledtabular}
      \begin{tabular}{lcc c c c c c}
         Collision system
                       & Data set & ${\langle}N_{part}{\rangle}$
                                        & $T_{ch}$ [MeV] &  $\mu_{q}$ [MeV] & $\mu_{s}$ [MeV] & $\gamma_s$ &  $\chi^{2}/N_{dof}$  \\
         \hline                                                                   
         130 GeV Au+Au & (1)      & 345 &   152.8$\pm$8.0 &   9.8$\pm$1.9  &   0.7$\pm$2.1   &   0.817$\pm$0.088  &  0.9/ 4 \\
                       &          & 289 &   152.4$\pm$8.1 &   9.7$\pm$1.9  &   0.7$\pm$2.2   &   0.911$\pm$0.098  &  0.5/ 4 \\
                       &          & 221 &   150.8$\pm$7.7 &   9.3$\pm$1.9  &   0.8$\pm$2.8   &   0.824$\pm$0.088  &  1.1/ 4 \\
                       &          & 132 &   151.1$\pm$7.7 &   8.8$\pm$1.8  &  -0.9$\pm$2.9   &   0.786$\pm$0.084  &  1.9/ 4 \\
                       &          & 102 &   150.2$\pm$7.4 &   8.3$\pm$1.8  &  -3.0$\pm$2.7   &   0.780$\pm$0.083  &  1.5/ 4 \\
                       &          &  48 &   147.3$\pm$6.9 &   7.3$\pm$2.0  &  -1.1$\pm$3.1   &   0.701$\pm$0.075  &  1.9/ 4 \\
                       & (2)      & 345 &   162.9$\pm$6.5 &  11.9$\pm$1.8  &   1.8$\pm$2.0   &   0.867$\pm$0.081  &  4.5/ 8 \\
                       &          & 289 &   160.7$\pm$6.2 &  11.2$\pm$1.7  &   1.3$\pm$2.1   &   0.958$\pm$0.090  &  2.9/ 8 \\
                       &          & 221 &   161.7$\pm$6.4 &  11.3$\pm$1.8  &   1.7$\pm$2.7   &   0.879$\pm$0.081  &  5.4/ 8 \\
                       &          & 132 &   161.9$\pm$6.3 &  10.8$\pm$1.8  &   0.0$\pm$2.8   &   0.840$\pm$0.077  &  6.3/ 8 \\
                       &          & 102 &   162.1$\pm$6.3 &  11.0$\pm$1.8  &  -1.3$\pm$2.7   &   0.837$\pm$0.076  &  7.6/ 8 \\
                       &          &  48 &   158.8$\pm$5.9 &   9.7$\pm$1.9  &   0.3$\pm$3.0   &   0.753$\pm$0.069  &  9.4/ 8 \\
                       & (3)      & 317 &   168.6$\pm$6.1 &  13.2$\pm$1.4  &   2.0$\pm$1.5   &   0.980$\pm$0.059  & 11.5/11 \\
                       &          & 211 &   165.6$\pm$5.9 &  11.9$\pm$1.8  &   1.7$\pm$2.5   &   0.930$\pm$0.060  &  8.4/11 \\
                       &          &  64 &   165.5$\pm$5.6 &  11.5$\pm$1.8  &   0.8$\pm$2.5   &   0.850$\pm$0.055  & 17.2/11 \\
                       & (5)      & 317 &   169.2$\pm$6.0 &  13.2$\pm$1.4  &   1.9$\pm$1.5   &   0.984$\pm$0.057  & 11.7/13 \\
                       & (6)      & 317 &   169.1$\pm$6.1 &  13.2$\pm$1.4  &   2.0$\pm$1.5   &   0.972$\pm$0.057  & 13.9/14 \\
         200 GeV Au+Au & (1)      & 351 &   152.4$\pm$3.3 &   7.9$\pm$2.1  &   1.5$\pm$4.6   &   0.963$\pm$0.063  &  1.2/ 3 \\
                       &          & 298 &   152.1$\pm$3.3 &   7.7$\pm$2.0  &   2.1$\pm$4.6   &   0.942$\pm$0.062  &  1.1/ 3 \\
                       &          & 234 &   151.6$\pm$3.2 &   7.4$\pm$2.0  &   0.3$\pm$4.6   &   0.927$\pm$0.060  &  1.4/ 3 \\
                       &          & 166 &   151.2$\pm$3.2 &   7.6$\pm$2.0  &   0.6$\pm$4.5   &   0.902$\pm$0.059  &  1.9/ 3 \\
                       &          & 115 &   151.2$\pm$3.2 &   7.3$\pm$2.0  &   0.6$\pm$4.5   &   0.884$\pm$0.058  &  2.3/ 3 \\
                       &          &  76 &   150.3$\pm$3.1 &   7.2$\pm$2.0  &  -0.5$\pm$4.5   &   0.853$\pm$0.056  &  3.1/ 3 \\
                       &          &  47 &   149.5$\pm$3.1 &   6.8$\pm$2.0  &  -0.6$\pm$4.5   &   0.822$\pm$0.054  &  4.9/ 3 \\
                       &          &  27 &   147.4$\pm$2.9 &   6.4$\pm$2.0  &  -1.3$\pm$4.5   &   0.776$\pm$0.051  &  7.2/ 3 \\
                       &          &  14 &   143.1$\pm$2.7 &   6.5$\pm$2.0  &   3.1$\pm$4.5   &   0.702$\pm$0.047  & 11.4/ 3 \\
                       & (2)      & 322 &   156.5$\pm$2.9 &   9.3$\pm$1.2  &   3.0$\pm$2.6   &   0.999$\pm$0.056  & 10.3/ 4 \\
                       &          & 234 &   156.5$\pm$3.5 &   9.6$\pm$2.4  &   0.0$\pm$4.7   &   0.987$\pm$0.059  & 13.1/ 4 \\
                       &          & 140 &   156.3$\pm$3.6 &   9.7$\pm$2.4  &   0.3$\pm$4.7   &   0.952$\pm$0.057  & 15.1/ 4 \\
                       &          &  76 &   155.3$\pm$3.5 &   9.5$\pm$2.4  &  -0.9$\pm$4.7   &   0.910$\pm$0.055  & 16.5/ 4 \\
                       &          &  29 &   152.4$\pm$3.3 &   8.7$\pm$2.3  &  -1.8$\pm$4.7   &   0.833$\pm$0.050  & 20.6/ 4 \\
                       & (3)      & 322 &   157.8$\pm$2.7 &   9.7$\pm$1.2  &   3.5$\pm$2.4   &   1.011$\pm$0.044  & 12.1/ 7 \\
                       &          & 234 &   158.5$\pm$3.2 &  10.9$\pm$2.2  &   2.3$\pm$3.8   &   0.990$\pm$0.046  & 14.4/ 7 \\
                       &          & 140 &   158.6$\pm$3.2 &  11.4$\pm$2.2  &   2.6$\pm$3.6   &   0.968$\pm$0.045  & 16.9/ 7 \\
                       &          &  76 &   157.2$\pm$3.0 &  10.8$\pm$2.1  &   0.9$\pm$3.6   &   0.920$\pm$0.042  & 17.4/ 7 \\
                       &          &  29 &   153.4$\pm$2.9 &   9.3$\pm$2.1  &  -1.5$\pm$3.8   &   0.850$\pm$0.043  & 21.1/ 7 \\
                       & (4)      & 322 &   156.3$\pm$2.7 &   9.4$\pm$1.2  &   3.4$\pm$2.4   &   1.020$\pm$0.045  & 19.5/ 8 \\
                       &          & 234 &   157.2$\pm$3.1 &  10.3$\pm$2.1  &   2.0$\pm$3.7   &   0.999$\pm$0.046  & 19.3/ 8 \\
                       &          & 140 &   157.8$\pm$3.0 &  10.9$\pm$2.1  &   2.5$\pm$3.6   &   0.965$\pm$0.044  & 18.6/ 8 \\
                       &          &  76 &   156.8$\pm$3.0 &  10.6$\pm$2.1  &   0.9$\pm$3.6   &   0.922$\pm$0.042  & 18.3/ 8 \\
                       &          &  29 &   153.2$\pm$2.8 &   9.2$\pm$2.1  &  -1.6$\pm$3.8   &   0.852$\pm$0.043  & 21.4/ 8 \\
                       & (5)      & 322 &   157.9$\pm$2.7 &   9.7$\pm$1.2  &   3.2$\pm$2.3   &   1.016$\pm$0.042  & 12.5/ 9 \\
                       & (6)      & 322 &   156.5$\pm$2.6 &   9.4$\pm$1.2  &   3.1$\pm$2.3   &   1.027$\pm$0.043  & 19.9/10
      \end{tabular}
    \end{ruledtabular}
  \end{table*}

  \begin{acknowledgments}
    We thank Jean Cleymans for useful discussion.
  \end{acknowledgments}

 \bibliography{chemical}

\end{document}